# The entire-sky catalog of isolated galaxies selected from 2MASS

© S.N. Mitronova [1], I.D. Karachentsev [1], V.E. Karachentseva [2], O.V. Melnyk [2]

1 Special Astrophysical Observatory of RAS, Nizhnij Arkhyz, Russia
2 Astronomical Observatory of Kiev National University, Kiev, Ukraine
E-mail: mit@sao.ru

**Abstract.** We present a first entire-sky catalog of isolated galaxies obtained via systematic automated and visual inspections of extended sources from the Two Micron All-Sky Survey (2MASS). Based on the Karachentseva's (1973) isolation criteria, we have extracted a sample of 2100 2MIG galaxies consisting a 4% fraction among galaxies brighter than $K = 12^m$. Being the objects with probably no major perturbations in some last billion years, the 2MIG galaxies may be considered as a reference sample to study environment effects in galaxy structure and evolution. The 2MIG catalog may also be useful to explore the existence of dark galaxy population within a volume of z=0.02.

## 1. Introduction.

The present observational data on galaxy distribution manifest the existence of cosmic large-scale web structure consisting of filaments, sheets and nodules (clusters) as their intersections. Such a fractal cellular structure may be naturally explained in the standard Lambda-CDM model. One can say roughly that only 5-10% of galaxies reside in dense (virialized) regions of cluster and about 10% others belong to immediate but unvirilized cluster surroundings. Most of the galaxies are situated in small systems like the Local group (50%) and loose clouds (25%). The remaining 5-10% of them are scattered in the general field. This sparse population of underdensed volumes (voids) seems to be an important tool to study the star formation history of galaxies subjected to very seldom external perturbations.

## 2. Finding algorithm for isolated galaxies.

A simple local method of identifying isolated galaxies was proposed by Karachentseva (1973). She selected all galaxies from Zwicky et al.(1961- 1968) catalog whose "significant" neighbors (of size within a factor four of the isolated galaxy candidate) lie further than twenty diameters away. It should be stressed here that application of the criterion requires knowledge of data on great number of galaxies much fainter than the isolated galaxy candidates. Karachentseva's (1973) catalogue (KIG) is composed of 1050 objects with apparent blue magnitudes brighter than $15.7^m$. The galaxies of KIG sample consist a small fraction, 3.6%, among the Zwicky galaxies. The KIG catalog was used by different authors to investigate the effects of under dense environment on galaxy properties (Adams et al. 1980, Haynes & Giovanelli, 1980, Verdes-Montenegro et al. 2005, and many others). A comprehensive studying the KIG galaxies is now going in frameworks of the AMIGA project (http://iaa.csic.es/AMIGA.html).

Karachentseva's (1973) criterion to find isolated galaxies was applied by us to the Extended Source Catalog of the Two Micron All-Sky Survey (XSC 2MASS) containing 1.64 millions of galaxies with K-band magnitudes brighter $14.5^m$. Under this criterion, a galaxy "i" with angular diameter $a_i$ is considered to be isolated if the projected sky separation $x_{in}$ between this galaxy and any neighboring galaxy "n" of angular diameter $a_n$ satisfies the following two relations:

$$x_{in} > s a_n \quad (1)$$

$$\frac{1}{4} a_i < a_n < 4 a_i \quad (2)$$

Here the dimensionless separation in Karachentseva's (1973) criterion was taken to be s=20. Under our a slightly modified version of the criterion (1)-(2), a galaxy "i" with a K-band magnitude, $K^{20fe}$, and isophotal K-band diameter $a^K = 2r^{20fe}$ is considered to be isolated if the projected sky separation between this galaxy and any neighboring one satisfies the same relations (1)-(2) with the same quantity s=20.

Applying this algorithm to find isolated objects, we considered only candidate galaxies with apparent magnitudes in the range of

$$4.0^m < K \leq 12.0^m \quad (3)$$

having angular diameters

$$a_K > 30". \quad (4)$$

The last two conditions are used to secure enough number of small neighboring galaxies around each isolated galaxy candidate. Note that XSC 2MASS catalog contains extended sources with $r^{20fe} > 5"$. This minimum galaxy size affects a bit the condition (2) when used together with (4).

Based on the conditions (1)-(4) with s=20, we identified 6218 isolated galaxies in the XSC 2MASS. The automated selection was carried out with the Pleinpot package environment designed for astronomical data reduction and analysis of the PostgreSQL Global Development Group.

**3. The 2MIG catalog.**

In Table 1 we present the basic properties of the 6218 isolated galaxies: Col.(1) a running identification number, Col.(2) the isolated galaxy name from 2MASS, Col.(3) the isolated galaxy radius $r^{20fe}$ in arcsec, Col.(4) the galaxy K-band magnitude $K^{20fe}$, Col.(5) the name of the most significant neighbor, Cols.(6),(7) the neighboring galaxy radius and magnitude, $r^{20fe}$ and $K^{20fe}$, Col.(8) the sky separation between the target and its neighbor in arcsec, Col.(9) the same but dimensionless separation, 2s, expressed in units of the angular radius of the significant neighbor. The total fraction of 2MIG isolated galaxies turns out to be 12%.

Table 1. Electronic catalog of 6218 isolated galaxies selected from 2MASS.

| N | Name1 | $a_{K1}/2$ | K1 | Name2 | $a_{K2}/2$ | K2 | x (") | 2s |
|---|---|---|---|---|---|---|---|---|
| 1 | 00001278+0107123 | 15.3 | 11.84 | 00000571+0115276 | 7.4 | 13.94 | 506 | 68.4 |
| 2 | 00002363-4701076 | 19.7 | 10.68 | 00001100-4707435 | 9.7 | 13.90 | 416 | 42.9 |
| 3 | 00002508+0751138 | 23.7 | 11.12 | 00000701+0816448 | 23.6 | 10.78 | 1554 | 65.8 |
| 4 | 00003295+0732582 | 16.1 | 11.45 | 00002508+0751138 | 23.7 | 11.12 | 1101 | 46.4 |
| 5 | 00005858-3336429 | 21.6 | 11.55 | 00020386-3328023 | 15.9 | 11.26 | 968 | 60.8 |
| 6 | 00010597-5359303 | 20.5 | 10.57 | 00003728-5359186 | 6.2 | 13.62 | 253 | 40.8 |
| 7 | 00011748-5300348 | 24.9 | 11.56 | 00020343-5259524 | 9.9 | 13.18 | 416 | 42.1 |
| 8 | 00013148+1120465 | 16 | 11.34 | 00011481+1118559 | 5.1 | 14.03 | 268 | 52.7 |
| 9 | 00013359+0900445 | 15 | 11.23 | 00010320+0900319 | 10.2 | 12.43 | 450 | 44.1 |
| 10 | 00013605-1444548 | 20.1 | 11.25 | 00015728-1441198 | 7.1 | 13.35 | 375 | 52.9 |
| | ... | ... | ... | ... | ... | ... | ... | ... |

As was shown by many authors, infrared diameters of galaxies in shallow 2MASS survey are systematically smaller than their standard optical diameters $a^{25}$. We found that the median ratio of optical-to-infrared diameters for KIG galaxies is $a^{25}/2r^{20fe} = 1.5$ with a large spread from one object to another. Taking into account this quantity, we selected a new sample of 2102 more isolated galaxies satisfying the same conditions (1)-(4), but with more rigorous quantity s = 30. All these very isolated galaxies were inspected visually to avoid cases when a galaxy isolatness may be destroyed by blue diffuse neighboring galaxies unseen in 2MASS. The visual inspection reduces the total fraction of very isolated galaxies from 4% to 3%, giving just the same relative number as in the KIG.

**4. Distributions of 2MIG galaxies in depth and in the sky.**

Figure 1 presents the integrated number vs. apparent K- magnitude diagram for the total XSC 2MASS sample (N= 1.64 millions), the XSC 2MASS galaxies with angular diameters $a^K$ greater than 30" (N=51572), isolated 2MIG galaxies (N=6218), and very isolated 2MIG galaxies (N=2102). Two solid lines correspond to the homogeneous distribution logN(<K) ~ 0.6K normalized at K=11.0$^m$. All the distributions demonstrate an excess of bright galaxies (K < 9$^m$) caused apparently by the Local supercluster. One can note that a fraction of isolated galaxies is almost the same for faint objects as well as for brighter ones. All-sky view of the new catalog of isolated galaxies is given in Figure 2 in equatorial coordinates. The upper map corresponds to the total 2MIG sample and the lower panel presents only very isolated galaxies (N=2102). As one can see, both the sky maps look to be quite homogeneous. There are no distinct signs of the Zone of Avoidance. Taken in comparison with the total 2MASS galaxy view (Jarrett, 2004), the 2MIG catalog finds only slight features of presence of the cosmic web structure.

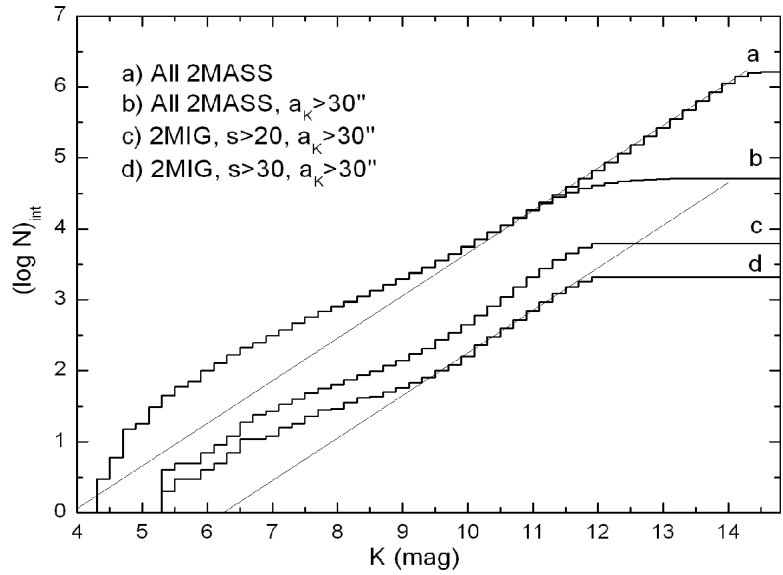

Fig.1. Integrated number of galaxies vs. apparent K-magnitude.

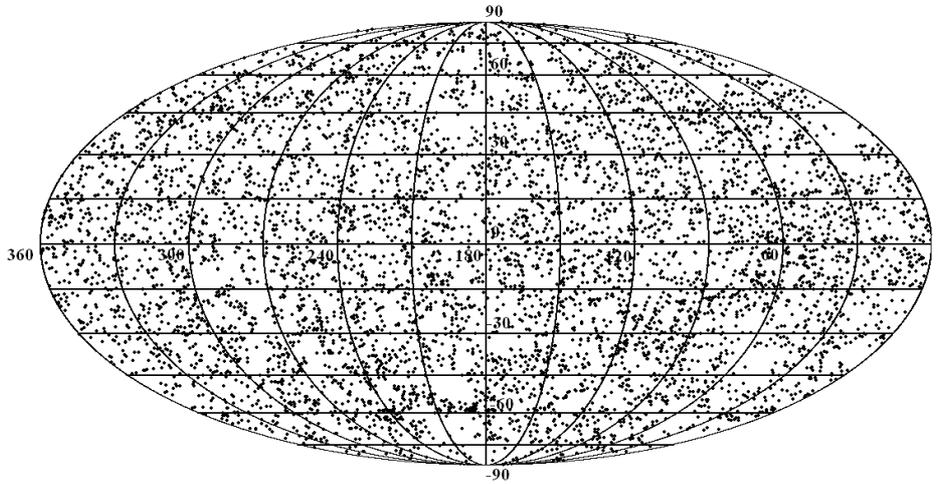

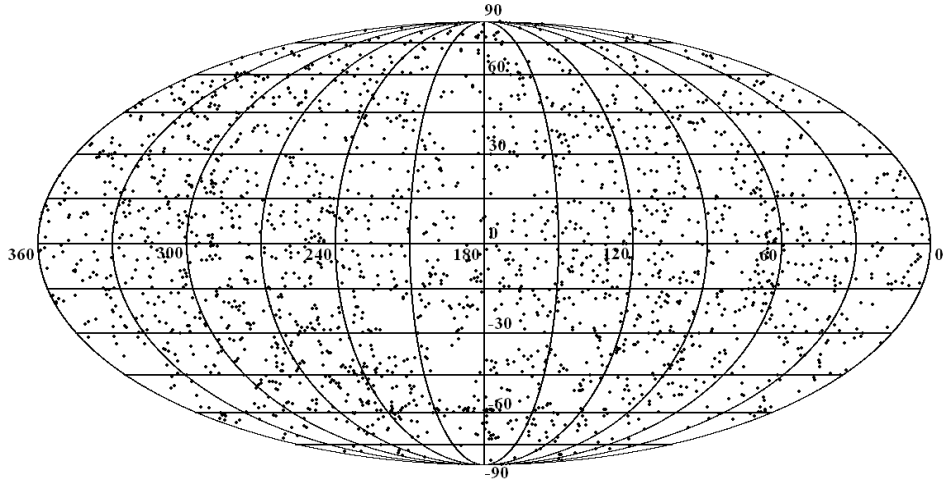

Fig.2. All-sky distribution of isolated 2MIG galaxies in equatorial coordinates.

**5. Conclusions.**

Preparing the new catalog of isolated galaxies, we did not take into account radial velocities of the galaxies. However, the majority of 2MIG objects have already measured radial velocities. The median redshift of our 2MIG sample reaches $z = 0.02$.

We carried out cross-identification of 2MIG galaxies with the KIG sample. About 30% of the KIG galaxies occur also in the new catalog. Such a rate of the catalog overlapping is a result of a bit different depth of two the samples as well as different optical-to-infrared diameter ratios for galaxies varying along Hubble sequence of morphological types. We note that 2MIG sample is the first entire-sky catalog of isolated galaxies that offers us an unique opportunity to explore the realm of galaxies habitant in low-density regions of the local universe.

Acknowledgements. We thank Dmitry Makarov for helpful discussions. This work was supported by the RFBR grant 07-02-00005 and DFG-RFBR grant 06-02-04017.